\documentclass[final,twocolumn,10pt]{elsarticle}
\usepackage{amssymb,amssymb,graphicx}
\usepackage{color}
\usepackage{graphicx}
\usepackage{booktabs}
\newcommand{\pn}{$\mathrm{P{_{N_2}}}$}
\newcommand{\p}{$\%$}

\newcommand{\Ei}{E$_\mathrm{i}$}

\newcommand{\R}{$ \rho $$_\mathrm{el}$}
\newcommand{\Esp}{$ < $E$_\mathrm{sp}$$ > $}
\newcommand{\Ts}{T$_\mathrm{s}$}
\newcommand{\Vb}{V$_\mathrm{b}$}

\begin{document}

\begin{frontmatter}

\title{Synthesis and study of highly dense and smooth TiN thin films}

\author{Susmita Chowdhury, Rachana Gupta, Shashi Prakash}

\address{Applied Science Department, Institute of Engineering and Technology, DAVV, Indore,
452017.}

\author{Layanta Behera, D. M. Phase and Mukul Gupta$^*$, }

\address{UGC-DAE Consortium for Scientific Research, University
Campus, Khandwa Road, Indore 452 001,India}

\address{$^*$ Corresponding author email: mgupta@csr.res.in}


\begin{abstract}

This study aims towards a systematic reciprocity of the tunable
synthesis parameters - partial pressure of N$_2$ gas, ion energy (\Ei) and Ti
interface in TiN thin film samples deposited using ion beam
sputtering at ambient temperature (300\,K). At the optimum
partial pressure of N$_2$ gas, samples were prepared with or without Ti interface
at \Ei~=~1.0 or 0.5\,keV. They were characterized using x-ray
reflectivity (XRR) to deduce thickness, roughness and density. The roughness of TiN thin films was found to be below 1\,nm, when deposited at the lower
\Ei~of 0.5\,keV and when interfaced with a layer of Ti. Under these
conditions, the density of TiN sample reaches to 5.80($\pm$0.03)\,g~cm$^{-3}$, a value highest hitherto for any TiN sample. X-ray diffraction and electrical resistivity measurements were performed. It was found that the cumulative effect of the reduction in
\Ei~from 1.0 to 0.5\,keV and the addition of Ti interface favors (111) oriented growth leading to dense and smooth TiN films and a substantial reduction in the electrical resistivity. The reduction in \Ei~has been attributed to the surface kinetics mechanism (simulated using SRIM) where the available energy of the sputtered species (\Esp) leaving the target at \Ei~= 0.5\,keV is the optimum value favoring the growth
of defects free homogeneously distributed films. 
The electronic structure of samples was probed using N K-edge absorption spectroscopy and the information about the
crystal field and spin-orbit splitting confirmed TiN phase
formation. In essence, through this work, we demonstrate the role
of \Esp~and Ti interface in achieving highly dense and smooth TiN thin films with low resistivity without the need of a high temperature or substrate biasing during the thin film deposition process.

\end{abstract}


\end{frontmatter}

\section{Introduction}
\label{1}

Titanium nitride (TiN) films are well-known for several applications~\cite{2020_TMN_Review,2003_MCP_TiN,2012_MCP_TiN,1998_MCP_TiN,2019_MCP_TiN}. They are propitious refractory
materials with their melting points exceeding 3000\,K~\cite{2020_TMN_Review,toth1971refractory} and in particularly have been used as diffusion barriers~\cite{huang2006effect}, alternative plasmonic materials ~\cite{patsalas2018conductive} in the visible and near-infrared (IR) regions at
both high~\cite{naik2012titanium,boltasseva2014empowering} and low
temperature~\cite{vertchenko2019cryogenic} regime and are well-known for providing
superhardness (=~28\,GPa)~\cite{pan1998mechanical} due formation of strong covalent
bonds between Ti and N
atoms~\cite{sundgren1986tinx}. Thin films of TiN can be synthesized in a large compositional range
within the NaCl type fcc structure and depending upon the
composition, the density can vary in a rather large
range~\cite{sundgren1986tinx}. Such variations in the ratio of Ti and N and the micro-structure leads to variable electrical resistivity of TiN films which typically varies between 12.4 to 500\,$ \mu $$ \Omega $ cm~\cite{karr2000effects,yokota2004resistivities}.

Very recently, Zhang et al.~\cite{zhang2019effects} investigated TiN thin films interfaced with Ti using a direct current magnetron sputtering (dcMS) process together with an applied  
substrate bias of -70\,V. Resulting films with (111) preferred 
orientation and low surface roughness showed the best
performance for a synchronizer ring required in automobile industry. In addition, the utilization of TiN as an electrode in
supercapacitors have  accelerated the microelectronics technology
in recent times~\cite{2020_TMN_Review}. Sun et al.~\cite{sun2019superior} have showed that TiN films deposited
at a substrate temperature (623\,K) and substrate bias (-150\,V
dc) lowered the leakage current density by three order of
magnitude compared to film deposited without substrate bias and
demonstrated that denser film (density = 5.41\,g\,cm$^{-3}$) with $\sigma$ $ < $ 1\,nm and low resistivity ($\approx$ 31\,$\mu$$\Omega$\,cm) plays a key role in the electrode
performance. Additionally, the reduction in
porosity of TiN films can also enhance the mechanical
behavior~\cite{patsalas2000effect,ma2006nanohardness}.
A study by Patsalas et al.~\cite{patsalas2000effect} has revealed that both hardness and
elastic modulus increases linearly with an increase in density of
TiN films and the highest density achieved was 5.7\,g\,cm$^{-3}$ for TiN film ($ \approx $ 100\,nm) deposited at a
substrate temperature of 673\,K and bias voltage of -100\,V, though the nominal value of density of TiN film is expected to be about 5.4\,g cm$
^{-3} $~\cite{patsalas2000effect}. However, due to large composition range and microstructure, the density of TiN films has been found to vary between 4.1 to 5.7\,g\,cm$
^{-3} $~\cite{patsalas2000effect,kearney2018substrate}. Moreover, the tunability of film
texturing (occurs due to lattice matched interface layers,
substrates or high adatom mobility)~\cite{patsalas2018conductive,jones2000effect,matias2015ion} and
compositional ratio (N/Ti atoms) can affect the hardness of the
film~\cite{huang2006effect,pan1998mechanical}. Several reports have confirmed that TiN films with (111) orientation comprises the highest
hardness~\cite{ljungcrantz1996nanoindentation,martinez2014effect,he2019effect} and serves as a template for (111)
oriented Al thin films, which reduces the
electromigration of Al metal ions useful in
microelectronics~\cite{kaizuka1994conformal}.

Here, it is to be noted that the addition of Ti (0002) as an interface escalates
the (111) texturing of TiN films due to the coherent atomic matching $ \approx
$\,98.4\,\% of both planes~\cite{ensinger1997low} and has been a topic of research for several decades. An alternate approach of such (111) texturing was demonstrated by Lattemann et al.~\cite{lattemann2010fully} utilizing high power impulse magnetron sputtering (HiPIMS), where the high-flux ionization species assisted the (111) grain growth even at room temperature with dense micro-structure. However, the dense morphology was confirmed by HRTEM measurements but the density of those TiN films ($ \approx $ 100\,nm) was not measured. Further, the HiPIMS process suffers from low deposition rates. Hence, it is clear that (111) textured films with improved
microstructural morphology (denser and smoother surface) is a
prerequisite from application based perspectives and few attempts
that were made are listed in
{Table~\ref{tab2}}.

Forepart, among the extensive adopted deposition techniques, relatively less attention has been dedicated for the synthesis of TiN thin films under ambient conditions i.e. without substrate bias and heating~\cite{cemin2019tuning,abadias2006interdependence,lattemann2010fully,yang2016room,merie2015research}. Deposition of dense TiN films under ambient conditions is not only cost effective, it also reduces the diffusion probability with the substrate or with other layers in a multilayers stack. The opted room-temperature deposition techniques were HiPIMS, dcMS or dual ion beam sputtering (DIBS) etc. where the sputtering powers, substrate temperatures, acceleration voltages, film thicknesses were modulated and even incorporation of Ti interface was done to detect the inter-dependence of the process parameters on the film growth and microstructures. On the contrary in this work, we used a simpler IBS process without any substrate heating or bias to deposit TiN thin films. The partial pressure of N${_{2}}$ gas was precisely controlled using a ultra-high vacuum (UHV) leak valve in the vicinity of the substrates. The incident ion energy (E$_\mathrm{i} $), was varied which in-turn varies the energy of the sputtered species $<$E$_\mathrm{sp}$$>$ leaving the target. Since, $<$E$ _\mathrm{sp} $$>$ is known to affect the thin film growth, its implication on the growth of TiN films deposited without or with Ti interface was studied. The self-hood of the work reclines in successful growth of denser and smoother TiN films by reducing the E$_\mathrm{i} $ from 1 to 0.5\,keV when interfaced with Ti, having the highest density of TiN thin film hitherto.  

\begin{table*} \center \caption{\label{tab2} A literature survey of TiN thin films deposited using different deposition methods and various process parameters without or with Ti interface. Here, dcMS = direct current magnetron
sputtering, rfMS = radio frequency magnetron sputtering, HiPIMS =
High power impulse magnetron sputtering, IBAD = ion beam assisted deposition, DIBS = dual ion beam sputtering, IBS = ion beam
sputtering, \Ts = substrate temperature, \Vb = substrate bias voltage, Ti = Titanium 
interface layer, $\sigma$ = surface roughness, \R~=~electrical resistivity.}
\begin{tabular} {llllll} \hline
Deposition 	&Process 	    &Density 	    	& $\sigma$ 	  & \R 		            & Reference \\
Method   &Parameters     &(g\,cm$^{-3}$)   & (nm)        &($\mu$$\Omega$\,cm)  &           \\
\hline
dcMS        &\Vb = -150\,V,\Ts = 623\,K &5.41  & $ < $ 1 & 31 &\cite{sun2019superior} \\
dcMS  & \Vb = 0\,V, \Ts = 623\,K & 4.77  & 1.9 & 105.2 &\cite{sun2019superior} \\
             dcMS & \Vb = -100\,V, \Ts = 673\,K & 5.7  & {--} & {--} & \cite{patsalas2000effect}    \\            
dcMS & \Vb $ > $ -100\,V; \Ts = 923\,K & 5.7  & {--} & 40 & \cite{patsalas2001optical}   \\
rfMS  & \Vb = 0\,V \Ts = 873\,K  & 4.14  & {--} & 70.2 & \cite{kearney2018substrate}\\
HiPIMS  & \Vb = 0\,V, \Ts  = 300\,K  & 5.30 & {--} & 130 & \cite{yang2016room}\\
IBAD  & \Vb = 0\,V, \Ts = 300\,K  & {--} & {--} & 400 & \cite{yokota2004resistivities}     \\
DIBS  & \Vb = 0\,V, \Ts = 300\,K  & {5.2~-~5.3} & {$ < $~1} & {--} &\cite{abadias2006interdependence}\\
dcMS & Ti, \Vb = -70\,V, \Ts  $<$ 423\,K  & {--}  & $ > $ 7.8 & {--} & \cite{zhang2019effects}  \\      
dcMS  & Ti, \Vb = 0\,V, \Ts = 723\,K  & {--} & {3.3} & {--} & \cite{chun1999dense}     \\                  
\textbf{IBS} & \textbf{\Vb = 0\,V, \Ts = 300\,K}  & \textbf{5.79} &  \textbf{0.7} &\textbf{117} & \textbf{this work}  \\
\textbf{IBS} & \textbf{Ti, \Vb = 0\,V, \Ts = 300\,K}  & \textbf{5.80}   & \textbf{0.6} & \textbf{91} &\textbf{this work } \\
            \hline
        \end{tabular}

\end{table*}

\section{Methodology}
\label{2} A series of TiN thin films without and with a Ti
interface were deposited on a single crystal Si (100), amorphous
glass and quartz substrates at room temperature (300\,K, no
intentional heating). Substrates were thoroughly cleaned in an
ultrasonic bath prior to the deposition. Samples were deposited
using a home-made radio frequency Ion Beam Sputtering (rf-IBS)
system~\cite{2011_JPCM_CuCo}. A Veeco 3\,cm rf (13.56\,MHz) ion
source was used to ionize 5N purity Ar gas flowing at 5\,sccm and
a similar amount of Ar gas was also used to neutralize the ion
beam. The Ar ion beam was kept incident at an angle of about
45$^{\circ}$ to sputter Ti target (purity 99.99\%). A schematic
diagram of the rf-IBS system is shown in the {Supplementary Material (SM)}~\cite{supplementaryinformation}. Prior to the
deposition, the base pressure achieved in the chamber was about
1$\times$10$^{-7}$\,Torr or lower. To achieve nitridation of
adatoms, N$_2$ gas was flown in the chamber in the vicinity of
substrates using a ultra-high vacuum (UHV) leak valve. The N$_2$
partial pressure (\pn) was raised from 1$\times$10$^{-7}$\,Torr to
2.5, 5.0 and 7.5 $\times$10$^{-5}$\,Torr. Incident Ar ion energy
E$_\mathrm{i}$ was kept either at 1.0 or at 0.5\,keV. Details of
samples and deposition conditions are given in
{table~\ref{tab1}}.

\begin{table} \caption{\label{tab1} Details of samples and deposition parameters used during deposition.
Here, \pn~is the nitrogen partial pressure and \Ei~is the incident
energy of the Ar$ ^{+} $ ions.}
\begin{tabular} {llll} \hline
Sample & Sample & \pn  & \Ei  \\
code & details & $\times10^{-5}$\  &   \\
& &(Torr)&(keV) \\ \hline
            Ti & Ti (10\,nm)  & 0.0   & 1.0      \\
            TiN-1 & TiN (100\,nm) & 2.5 & 1.0      \\
            TiN-2 & TiN (100\,nm) & 5.0  & 1.0      \\
            TiN-3 & TiN (100\,nm) & 7.5  & 1.0      \\
            TiN-4 & TiN (100\,nm) & 2.5  & 0.5      \\
            TiN-5 & Ti(10\,nm)/TiN(90\,nm)  & 2.5  & 1.0      \\
            TiN-6 & Ti(10\,nm)/TiN(90\,nm)  & 2.5  & 0.5      \\
           \hline
        \end{tabular}
        \end{table}

The long-range structural ordering was
characterized by x-ray diffraction (XRD) using Bruker D8 Advance
XRD system based on $\theta$-2$\theta$ Bragg-Brentano geometry.
The Cu-K$\alpha$ source produced x-rays of wavelength 1.54\,{\AA}
and they were detected using a fast 1D detector (Bruker LynxEye).
The density, thickness and roughness of the as-deposited samples
were analyzed using x-ray reflectivity (XRR) measurements
performed on a Bruker D8 Discover system using Cu-K$\alpha$
x-rays. To probe the surface morphology at the atomic scale level,
atomic force microscopy (AFM) measurements were done in
non-contact mode over a scan area of 2$\times$2 $ \mu $m$^{2}$ (can be seen in SM)~\cite{supplementaryinformation}. The dc electrical resistivity {\R} of the samples were performed by collinear standard four-probe method at room temperature keeping the current fixed at $ \pm $~105\,$ \mu $A. 
Surface sensitive soft x-ray absorption near edge spectroscopy
(SXAS) measurements were carried out at BL-01~\cite{2014_AIP_BL01}
beamline at Indus-2 synchrotron radiation source at RRCAT, Indore,
India. The synchrotron radiation source was operating at an
electron energy of 2.5\,GeV and a beam current of 150\,mA. SXAS
measurements were performed at N K-edge and Ti L$_{3,2}$ edges in
total electron yield (TEY) mode which allows to collect all the
electrons ejected out from the sample surface probing a depth of
about 10\,nm~\cite{henderson2014x}. A computational analysis of
$ < $E$_\mathrm{sp}$$ > $ was performed by stopping and range of ions in
matter (SRIM)~\cite{ziegler2010srim}. The analysis was performed for a
sampling of 10$^{5}$ Ar$^{+}$ ions impinging on the target surface
with an angle of inclination of 45$^{\circ}$, similar to the
experimental condition used in this work.

\section{Results and discussion}
\label{3}
\subsection{X-ray reflectivity}

\begin{figure}
  \includegraphics
    [width=0.6\textwidth] {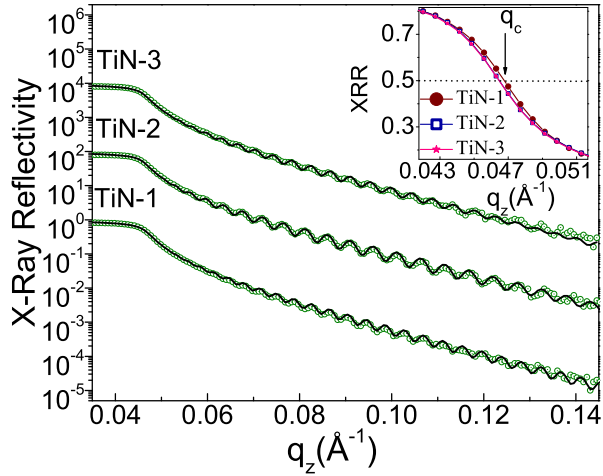}
  \vspace{-10mm}
    \caption{X-ray reflectivity pattern of TiN samples deposited at various \pn of 2.5$\times$10$ ^{-5} $ (TiN-1), 5$\times$10$ ^{-5} $ (TiN-2) and 7.5$\times$10$ ^{-5} $\,Torr (TiN-3). The patterns are shifted along the y-axis for clarity. The inset compares the variation in the vicinity of critical angle region. Here, the open circles represent the experimental data and solid lines show the theoretical fitting.}
  \label{Fig:1}
\end{figure}

The XRR pattern of TiN thin films prepared using different values
of $\mathrm{P{_{N_2}}}$ at 1\,keV (TiN-1, TiN-2 and TiN-3) are
shown in {Fig.~\ref{Fig:1}} as a function of momentum
transfer vector (q$ _{z}$). They were fitted using Parratt 32
software~\cite{braun1999parratt32} based on Parratt's
formalism~\cite{PhysRev.95.359}. The inset of
{Fig.~\ref{Fig:1}} compares XRR data of these samples
depicting small changes in the critical q$ _{z}$ (taken
approximately at 50\p~intensity) given by q$_{c}$ = 4$
\pi$sin$\theta_{c}$$ / $$ \lambda $, where $\theta_{c}$ =
$\sqrt{2\delta}$ = $ \sqrt{\lambda^{2}\rho/\pi} $ is the critical
angle, $\delta$ is dispersive part of the refractive index of the
material, $\rho$ is scattering length density (SLD) of the
material and $\lambda$ is the wave-length of
x-rays~\cite{tolan1999reflectivity}. With the known values of $\rho$ obtained
from the fitting of XRR data, mass density ($\rho$$_\mathrm{m}$)
of samples were obtained (assuming TiN stoichiometry). The fitting
step includes a thin ($ < $ 3\,nm) titanium oxynitride layer (TiO$
_{x} $N$ _{y} $) on TiN thin films followed by an interface layer
($ < $ 2\,nm) of some titanium oxide (Ti$ _{x} $O$ _{y} $) on Si
substrates. It is known that TiN films are affected by surface
oxidation to a large extent on exposure to atmosphere due to
higher affinity of Ti towards oxygen ($ \Delta $H$ _{f}^{0} $
ranging between -542 to -1521\,kJ mol$ ^{-1} $ for different
compositions of Ti$ _{x} $O$ _{y} $ at 298K). Howbeit, in our TiN
thin films the SLD of the topmost oxide layer comes out to be
higher than the known forms of Ti$ _{x} $O$ _{y} $ indicating the
presence of some TiO$ _{x} $N$ _{y} $. Here it is worth mentioning
that with increase in \pn, the contribution of surface oxidation
gradually reduces, as expected. Moreover, even the presence of
negligible amount of O$ _{2} $ in the deposition chamber can also
form oxide layer in the substrate-film interface which was also
evidenced in our samples. The O$ _{2} $ base pressure was $
\approx $ 7.5$ \times $10$ ^{-10} $\,Torr in our chamber (measured
using a residual gas analyzer)~\cite{tiwari2019interface}, and it
seems that this negligible amount of O$ _{2} $ readily reacts with
Ti forming a thin ($ < $ 2\,nm) Ti$_{2}$O$ _{3} $ phase during the
early stage of growth ($ \Delta $H$ _{f}^{0} $ = -1521\,kJ mol$
^{-1} $ for Ti$_{2}$O$ _{3}$ compared to -337\,kJ mol$ ^{-1} $ for
TiN). Detailed fitted parameters are listed in
{Table~\ref{tab3}}.

The deposition rate (DR) using pure Ar was 40.2\,\AA/min for Ti,
and decreases to 21.9, 17.8, and 16.1\,\AA/min in TiN-1, TiN-2 and
TiN-3 samples. A reduction in DR can be associated to the
nitridation of the target itself known as target
poisoning~\cite{SCHILLER1984259,BERG2005215,anders2017tutorial}.
Generally, DR from a fully poisoned target become negligibly
small~\cite{stanislav1990properties,sundgren1983mechanisms}. In the
present case, the TiN-3 sample (\pn = 7.5$\times$10$ ^{-5} $ Torr)
seems to be at the onset of target poisoning but the formation of
a TiN phase in TiN-1 and TiN-2 samples is taking place from a so
called transient state lying somewhere between the metallic and
poisoned state~\cite{sundgren1983mechanisms}. Such a reduction in
DR can be attributed to the formation of much stronger chemical
bonds of the compound on the target surface (with the introduction
of N$_{2}$ flow) than the elemental metallic bonds present within
the target resulting in lowering of DR~\cite{anders2017tutorial}.
Subsequently, the SLD also reduces as can be seen from
{Table~\ref{tab3}} yielding reduction in
$\rho$$_\mathrm{m}$ from 5.42 (TiN-1) to 5.36 (TiN-2) and to
5.33\,g cm$ ^{-3} $ (TiN-3). From the observed values, it can be
seen that the density of the sample labelled as TiN-1 appears to
be in the closest agreement with that of stoichiometric bulk TiN
($ \approx $ 5.4\,g cm$ ^{-3} $)~\cite{patsalas2000effect}. Within the
present experimental conditions (using a UHV leak valve), the
minimum control level of \pn was about 5$\times$10$^{-6}$\,Torr
and already at a low partial gas flow of 2.5$\times$10$^{-5}$\,Torr, stoichiometric TiN is
achieved. When flow is increased further (which was also evidenced by the residual gas analyzer), it apparently creates
voids or disorder leading to a reduction in the density of the
films. This becomes clearer after comparing the XRD results
presented in section {3.2}.

           \begin{table} \caption{\label{tab3} Parameters of TiN thin films obtained from of x-ray reflectivity (XRR) measurements. Here, scattering length density (SLD) and rms surface roughness ($\sigma$) was obtained from fitting of XRR data. }
        \begin{tabular} {lll} \hline
            Sample & SLD & $ \sigma $   \\
            code & ( $ \pm $ 0.02 &  \\
            & $\times$ 10$ ^{-5} $\,\AA$ ^{-2} $ ) &($ \pm $ 0.2\,nm)\\ \hline
           TiN-1 & 4.35  & 1.4 \\
            TiN-2 & 4.30  & 1.3  \\
            TiN-3 & 4.28  & 1.4  \\
           TiN-4 & 4.65  & 0.9 \\
           TiN-5 & 4.60  & 0.9 \\
           TiN-6 & 4.66  & 0.9 \\
            \bottomrule
       \end{tabular}
    \end{table}

\subsection{X-ray diffraction}

\begin{figure}
    \includegraphics[width=0.6\textwidth] {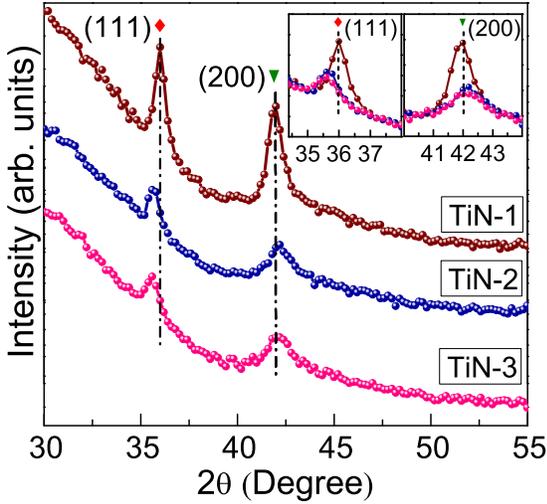} \vspace{-10mm}
    \caption{X-ray diffraction pattern of TiN samples deposited at ambient temperature (300K) for different \pn of 2.5$\times$10$ ^{-5} $ (TiN-1), 5$\times$10$ ^{-5} $ (TiN-2) and 7.5$\times$10$ ^{-5} $\,Torr (TiN-3). The inset compares the peak positions around (111) and (200) peaks.}
    \label{Fig:2}
\end{figure}

{Fig.~\ref{Fig:2}} shows the XRD patterns of TiN-1, TiN-2
and TiN-3 samples. In case of TiN-1 sample, prominent peaks
occurring at 2$\theta$ = 36.08 and 42.05$^{\circ}$ can be assigned
to TiN (111) and TiN (200) reflections~\cite{patsalas2004surface}.
With increasing \pn, the peak corresponding to (111) lattice plane
shifts towards lower 2$ \theta $ whereas the (200) peak slightly
displaces towards the higher angle region (depicted more clearly
in the inset of {Fig.~\ref{Fig:2}} for TiN-1, TiN-2 and
TiN-3 samples). However, both these reflections become broadened
due to lack of crystalline long range ordering with increasing
\pn. Detailed values of lattice parameters (LP) obtained from both
lattice planes and the crystallite size (D) obtained from the most
intense (111) peak using Scherrer's formula are given in
{SM}~\cite{supplementaryinformation}. For TiN-1, the
deduced LP are a$^{111} $ = 4.30\,($\pm $ 0.01)\,{\AA} and
a$^{200} $ = 4.29\, ($\pm$ 0.01)\,{\AA}. Similar values (within
the experimental accuracy) of LP obtained from both the (111) and
(200) lattice planes are expected in a cubic symmetry. Comparing
the value of LP in our samples, it is evident that the samples exhibit in-plane compressive stress common in room temperature deposited ion beam sputtering techniques~\cite{abadias2006interdependence} but lies within the experimental
values obtained in different
works~\cite{patsalas2000effect,PELLEG1991117,esaka1997comparison,VALVODA1988225}. Generally, the LP of TiN has been found anywhere between 4.18 -
4.40\,{\AA}~\cite{PELLEG1991117,esaka1997comparison,sundgren1985structure},
where in a number of works the LP of 4.30\,{\AA} has also been
obtained and assigned to be in $ \delta $-TiN
phase~\cite{patsalas2000effect,PELLEG1991117,esaka1997comparison,VALVODA1988225}.
Therefore, it can be concluded that our TiN-1 sample is closest to
stoichiometric $ \delta $-TiN, which is also supported by XRR
results.

By increasing the \pn, we find that LP obtained from the (111)
reflection increases whereas that of (200) decreases. This
indicates some distortion from the cubic symmetry
~\cite{elstner1994structure}. As we find from our XRR results also,
the density of film decreases in TiN-2 and TiN-3 samples. It
appears that over-saturation of nitrogen in TiN may lead to such
effects.

\subsection{Influence of ion energy on growth behavior of TiN thin films}

\begin{figure}
\includegraphics[width=0.62\textwidth] {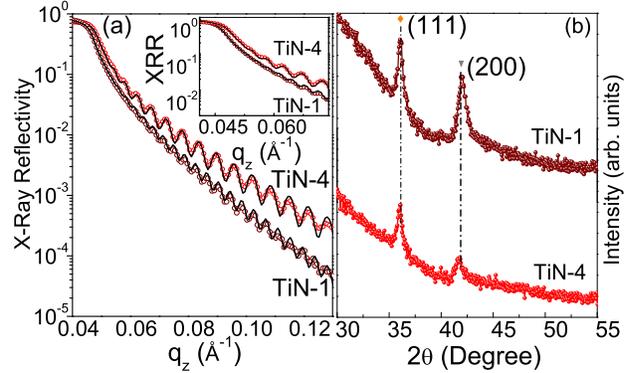} \vspace{-10mm}
\caption{XRR (a) and XRD (b) patterns of TiN-1 and TiN-4 thin films deposited at variable incident ion energies of 1 and 0.5\,keV with a constant Ar gas flow of 5\,sccm and \pn of 2.5$\times$10$ ^{-5} $\,Torr. The inset of  (a) compares the critical density region of TiN-1 and TiN-4 thin films. Here, tue open circles represent the experimental data and solid lines show the theoretical fitting in (a).}
\label{Fig:3}
\end{figure}

From the results shown in section {3.1} and {3.2}
(XRR and XRD results), we found that nearly stoichiometric TiN
phase was obtained in TiN-1 sample. Therefore, keeping the
parameters similar as in TiN-1, a sample was grown by reducing E$
_\mathrm{i} $ to 0.5\,keV and labelled as TiN-4. A comparison of
XRR and XRD data of TiN-1 and TiN-4 samples are shown in
{Fig.~\ref{Fig:3}}. The fitting of XRR pattern of TiN-4
sample was also performed following a similar procedure as
mentioned in section {3.1}. Here the data of TiN-1 sample
is the same as presented in {Fig.~\ref{Fig:1}} and
{\ref{Fig:2}}. The effect of reduction of E$ _\mathrm{i} $
on the XRR pattern can be seen in terms of a less steeper decay of
XRR pattern. It is known that XRR pattern is affected by the rms
roughness ($\sigma$) of the sample surface as, R $\propto$ e$
^{-q_{z}^{2}\sigma^{2}} $ where R is the specular reflectivity
from the sample surface~\cite{gibaud2000x}. Therefore, it becomes
clear that the roughness of the sample deposited at lower E$
_\mathrm{i} $ is smaller. In addition, the SLD of this sample also
comes out to be higher (this becomes evident by comparing the q$
_{c} $ region as shown in the inset of {Fig.~\ref{Fig:3}
(a)} yielding $\rho_\mathrm{m}$ = 5.79\,g cm$ ^{-3} $. Fitted
parameters of TiN-4 sample are also shown in
{Table~\ref{tab3}}. It is somewhat surprising to note that
the density of TiN-4 sample at 5.79\,g cm$ ^{-3} $ surpasses the
previous highest value of 5.7\,g cm$ ^{-3} $ achieved by Patsalas
et al.~\cite{patsalas2000effect} in TiN thin film samples grown at a
very high substrate temperature (673\,K) together with an applied
bias voltage of -100\,V. In contrast, our sample was grown at room
temperature (300\,K) without any substrate biasing. From the
fitting parameters as shown in {Table~\ref{tab3}}, we can
see that roughness ($ \approx $ 0.9\,nm) of the film reduces to
sub-nanometer region inspite of the fact that the film thickness
being fairly large at about 90\,nm.

The effect of reduced E$ _\mathrm{i} $ on the XRD pattern is shown
in {Fig.~\ref{Fig:3} (b)}. In comparison to the TiN-1
sample, we can see that the peak positions in this sample (TiN-4)
are also similar resulting in similar value of
LP~\cite{supplementaryinformation}. On the other hand, the
significant difference between TiN-1 and TiN-4 samples can be seen
in terms of relative intensities of (111) and (200) reflections as
can be observed from the XRD fitting. In case of TiN-1, I$
_{(111)} $ = 0.57 I$ _{(200)} $ signifying that (200) reflection
is more pronounced compared to (111), where I$ _\mathrm{(hkl)} $
signifies the area under the curve for the particular (hkl)
reflection whereas for TiN-4, I$ _{(111)} $ = 1.79 I$ _{(200)} $
clearly stating that texturing of (111) reflection is favored in
the later case. Moreover, the prominent effect in the reduction of
E$ _\mathrm{i} $ can be manifested in terms of reduced roughness
and increasing density and it can be correlated to {\Esp} of the 
sputtered species leaving the target which in turn modifies the growth kinetics by changing the adatom diffusion lengths during the film growth. To evaluate {\Esp}, SRIM simulations
were performed on Ti targets in pure Ar environment and the effect of surface binding energy was also taken into account to obtain the correct values~\cite{ziegler2010srim,mahieu2009reactive}. For E$
_\mathrm{i} $~=~1\,keV and 0.5\,keV, the calculated values of {\Esp} (eV/atom) are 35.2$ \pm $2 and 27.1$ \pm $2,
respectively which can be suitably translated to TiN. The effect
of such influence on {\Esp} ($ \approx $ 20\p\,reduction by
changing E$ _\mathrm{i} $ from 1 to 0.5\,keV) is in particularly
significant for TiN whereas other elements (e.g. nickel) does not
seem to be affected so much by such an incremental change in
E$_\mathrm{i}$. This was amply demonstrated on surfactant mediated
growth of Ti/Ni multilayers~\cite{gupta2011surfactant}. The
consequence of variation in adatom mobility is enhanced (200)
reflection occupying the lowest surface energy site in TiN-1,
albeit in TiN-4 the preferred orientation is reversed (along (111)
plane) favoring the closely packed plane by lowering the overall
energy of the crystal which is expected in kinetics driven
mechanism~\cite{patsalas2000effect}.

In essence, the effect of reduction in the E$ _\mathrm{i} $ has
positively influenced the growth of TiN film which has a very
dense microstructure as well as low roughness. However, since the
I$_\mathrm{(200)}$ has not been completely suppressed, we utilized
a Ti interface both at E$ _\mathrm{i} $~=~1.0 and 0.5\,keV in the
next section {3.4}.

\subsection{Effect of Ti interface on texturing of TiN thin films}

\begin{figure}
    \includegraphics[width=0.6\textwidth] {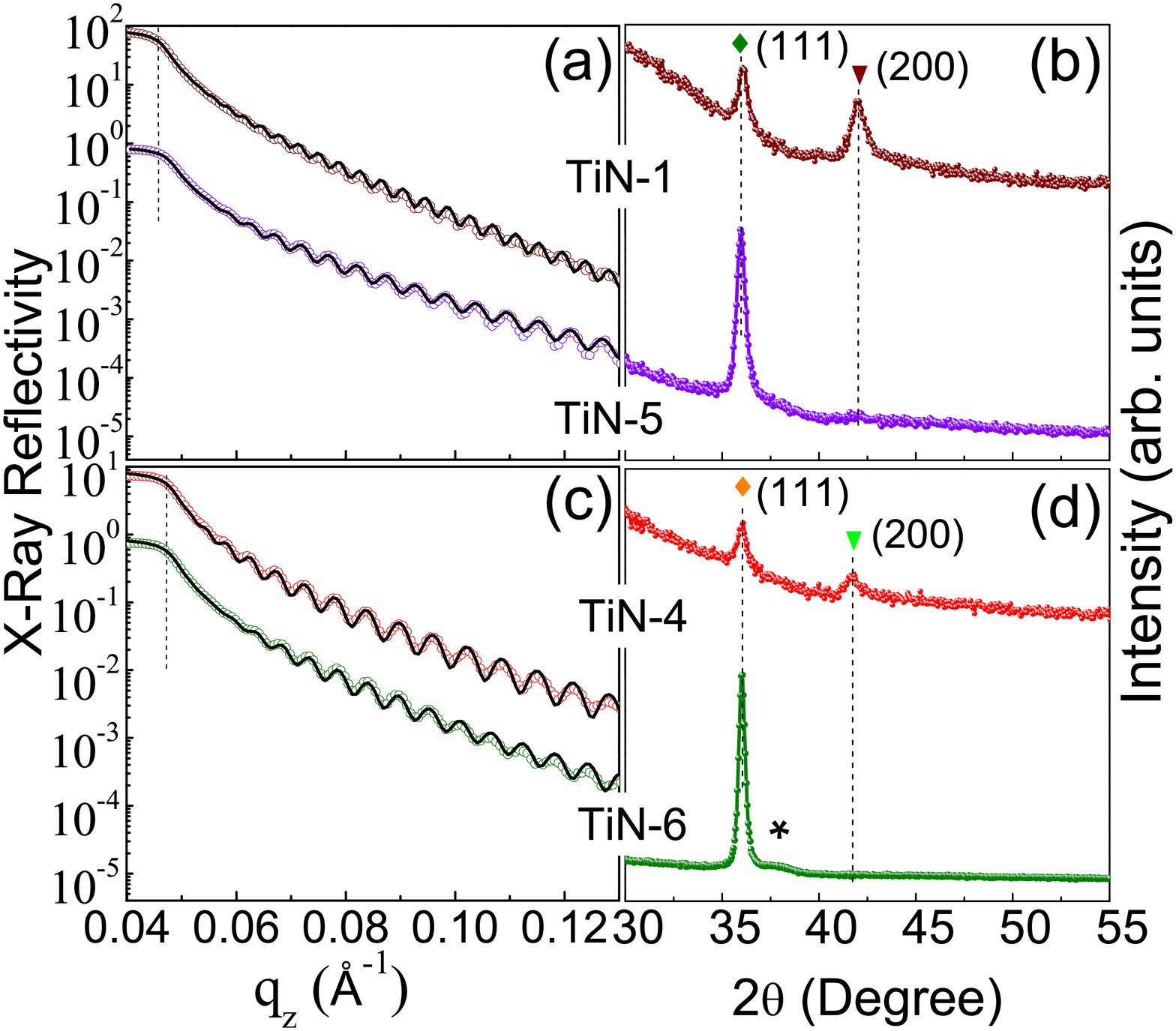} \vspace{-10mm}
    \caption{The XRR (a) and XRD (b) data of TiN-1 and TiN-5 samples deposited at an ion energy of 1\,keV with same Ar gas flow pf 5\,sccm and \pn of 2.5$\times$10$^{-5}$\,Torr. A comparison of XRR (c) and XRD (d) patterns are shown at a reduced ion energy of 0.5\,keV for TiN-4 and TiN-6 thin films deposited at same deposition parameters. Here, the open circles represent the experimental data and solid lines show the theoretical fitting in (a) and (c). }
    \label{Fig:4}
\end{figure}

{Fig.~\ref{Fig:4} (a)} and {\ref{Fig:4} (b)}
compares the XRR and XRD data of samples prepared without (TiN-1)
and with Ti interface (TiN-5) deposited under similar conditions
(\pn of 2.5$\times$10$ ^{-5} $\,Torr, E$ _\mathrm{i} $~=~1.0\,keV)
whereas {Fig.~\ref{Fig:4} (c)} and {\ref{Fig:4}
(d)} compare the samples deposited without (TiN-4) and with Ti
interface (TiN-6) deposited at reduced E$ _\mathrm{i} $ = 0.5\,keV
keeping \pn unaltered. As mentioned earlier in section
{3.1} and {3.3}, a thin layer of TiO$ _{x}
$N$ _{y} $ phase was introduced for TiN-5 and TiN-6 samples as
well, corresponding to surface oxidation. The same intermediate
phase of Ti$ _{x} $O$ _{y} $ ($ < $ 2\,nm) was also introduced
between Ti and Si surface for TiN-5 and TiN-6 samples. The fitted
parameters are shown in {table~\ref{tab3}}. The $
\rho_\mathrm{m} $ values obtained from SLD which was initially
5.42\,g cm$^{-3}$ for TiN-1 film increased to 5.73\,g cm$^{-3}$
for TiN-5 sample with the introduction of Ti interface. This is
expected because the close packed hexagonal structure of Ti
interface promotes growth of smoother and denser TiN films with a
low surface roughness of 0.9\,nm. Moreover, mixed orientations
found in TiN-1 sample get completely diminished and texturing of
(111) plane was achieved with a small trace of (200) reflection in
the presence of thin (10\,nm) Ti interface for TiN-5 film (90\,nm)
as shown in {Fig.~\ref{Fig:4} (b)}. It is well known that such unidirectional
grain growth occurs due to the coherent atomic matching
($\approx$\,98.4\,\%) of the (0002) Ti plane with the (111) plane
of TiN leading to minimization of the interfacial
strain~\cite{jones2000effect,chun1999dense}. Additionally, Ti interface provides
better adhesion between film and the substrate by reducing the
residual stress~\cite{larsson1995deposition}.

Similar trends in XRR and XRD were also observed for TiN-4 and
TiN-6 samples (as shown in {Fig.~\ref{Fig:4} (c)} and
{\ref{Fig:4} (d)}). The $ \rho_\mathrm{m} $ of TiN-6 film
comes out to be 5.8\,g cm$ ^{-3} $, even larger than that of TiN-4
sample and the roughness remains low at 0.9\,nm in both samples.
From the XRD pattern as shown in {Fig.~\ref{Fig:4} (d)},
the presence of Ti phase (the broad feature shown by an asterisk ($ \ast $)) was
detected accompanied with a highly textured (111) orientation of
the TiN-6 film. In addition, the (200) plane was completely
suppressed for TiN-6 films in comparison to TiN-4. Hence, from the
above results it is evident that kinetics of lowering in adatom
mobility (due to variation of {\Esp} as obtained from SRIM calculation) consequences in
preferred orientation of densely packed (111) plane with
relatively smaller grain size (as can be seen from
{SM}~\cite{supplementaryinformation}) which reduces defects
(in the form of voids) in the crystal lattice and results in
densification as seen for TiN-4 thin films and the addition of Ti
interface results in texturing of the TiN-6 film with dense
microstructure. Moreover, increase in density of TiN-6 films with
addition of densely packed Ti interface is less pronounced in the
present case as the film was already voids free during low ion
energy deposition (as for TiN-4) but it promoted larger grain growth of the TiN-6 sample.

\subsection{Room-temperature resistivity}

\begin{figure}
	\centering
	\includegraphics[width=0.4\textwidth] {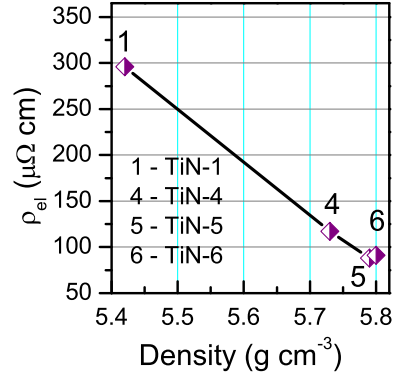}
	\vspace{-10mm}
	\caption{The variation in the electrical resistivity (\R) of TiN thin film samples deposited at {\Ei}~=~1.0 (TiN-1 and TiN-5) and 0.5\,keV (TiN-4 and TiN-6) without (TiN-1 and TiN-4) and with (TiN-5 and TiN-6) Ti interface at a constant Ar gas flow of 5\,sccm and \pn = 2.5$\times$10$ ^{-5} $\,Torr.}
	\label{Fig:6}
\end{figure}

Furthermore, to get an insight on the electrical transport behavior, the resistivity (\R) measurements were performed at room-temperature. The variation of \R~with the density of different TiN thin film samples is shown in Fig.~\ref{Fig:6}. As can be seen from this figure that \R~decreases almost linearly moving from sample TiN-1 to TiN-4 to TiN-5/TiN-6. This implies that the reduction of \Ei~from 1 to 0.5\,keV or the introduction of Ti interface or both lead to denser films with lower values of \R. Here, it is worth mentioning that the \R~of TiN typically lies between 30 to 450\,$\mu$$\Omega$ cm ~\cite{yokota2004resistivities,cemin2019tuning} but application of substrate bias, substrate temperature etc. substantially reduces the \R~values due to annihilation of the defects and structural relaxation at relatively higher energy~\cite{yokota2004resistivities,kearney2018substrate}. For room temperature deposited TiN films, Yang et al. reported a value of 130\,$\mu$$\Omega$ cm using HiPIMS~\cite{yang2016room}. In contrast, \R~=~400\,$\mu$$\Omega$ cm was obtained by Yokota et al. in IBAD~\cite{yokota2004resistivities} deposited TiN thin films. And even \R~{$>$}~500\,$\mu\Omega$ cm was reported by Kearney et al.~\cite{kearney2018substrate} using rfMS. Fig.~\ref{Fig:6} depicts an inverse relation between the \R~and density of the samples, which is a general trend that is indeed expected~\cite{cemin2019tuning}. Subsequent to the fact that all the samples typically exhibit the same thickness (90~-~100\,nm) and have been deposited at same temperature (300\,K), such a variation can be attributed to the suppression of the defects across the film surface and the grain boundary sites due to significant reduction in $<$E$_\mathrm{sp} $$>$ and addition of Ti interface, which in-turn leads to larger grain growth and considerable reduction in porosity of the films. It yields lowering in scattering of the conduction electrons with the defects, which plays a dominant role at ambient temperature~\cite{roy2019quantum,ningthoujam2015synthesis}. The result is in accordance with the XRD data as well, where an inverse proportionality between the crystallite size and the resistivity of the samples can be observed (see SM)~\cite{supplementaryinformation}.

\subsection{Soft X-ray Absorption Spectroscopy}

\begin{figure}
      \includegraphics[width=0.7\textwidth]  {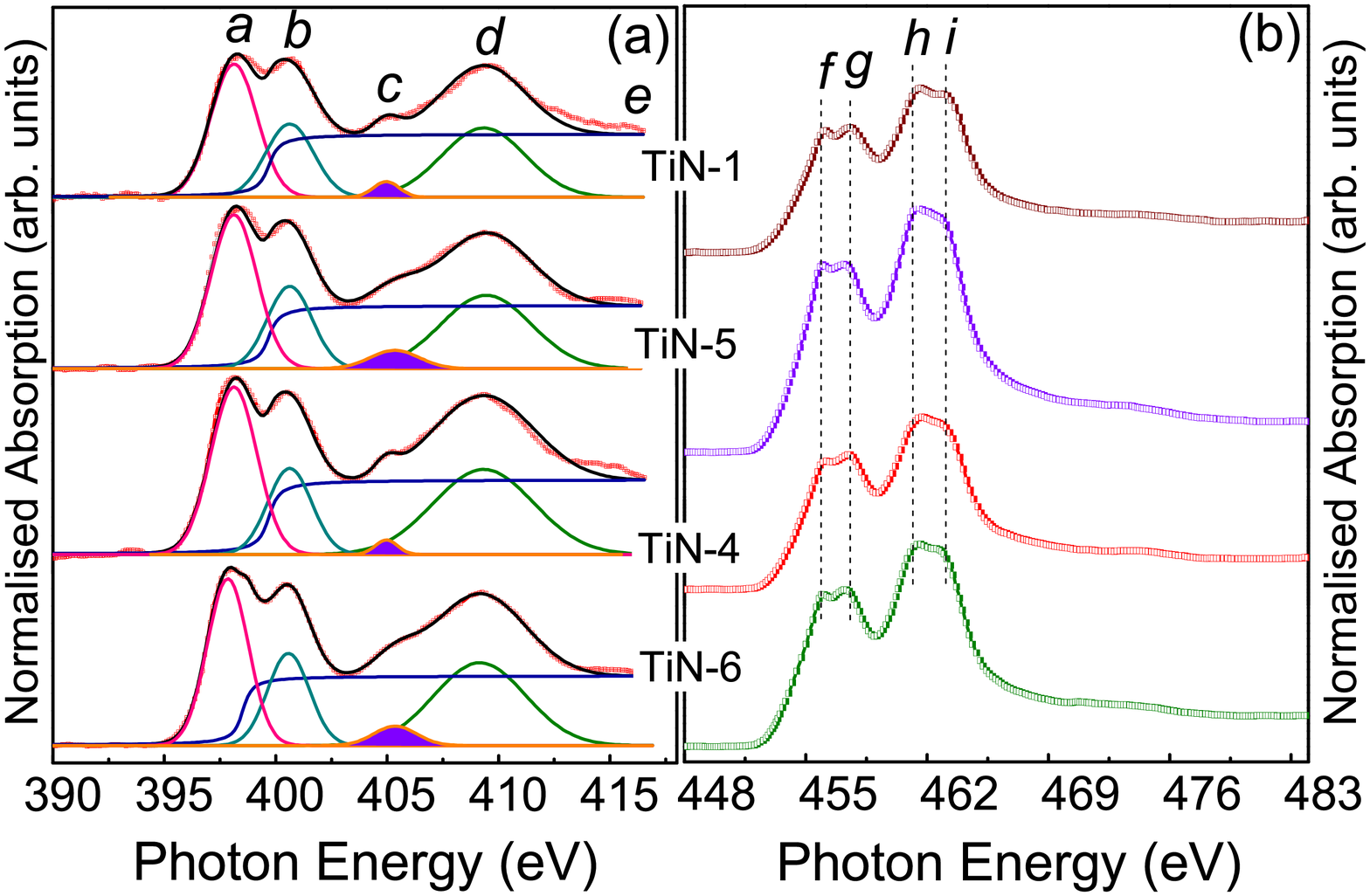} \vspace{-10mm}
    \caption{Normalised SXAS spectra of fitted N K-edge (a) and Ti L-edge (b) of TiN thin films deposited without and with Ti interface at variable incident ion energies of 1 and 0.5\,keV at a constant Ar gas flow of 5\,sccm and \pn = 2.5$\times$10$ ^{-5} $\,Torr. Here, the open circles represent the experimental data and solid lines show the theoretical fitting in (a).}
    \label{Fig:5}
\end{figure}


{Fig.~\ref{Fig:5} (a)} compares the normalised and fitted
N K-edge spectra whereas {Fig.~\ref{Fig:5} (b)} shows the
Ti L$ _{3,2} $ features of TiN-1, TiN-4 (E$ _\mathrm{i} $ =
1\,keV) and TiN-5, TiN-6 (E$ _\mathrm{i} $ = 0.5\,keV) samples
deposited without and with Ti interface. It is well known that
intensity of XAS spectra probe the unoccupied density of states
making it an element specific technique. The pre and post edge
normalisation as well as the fitting of N K-edge spectra were
processed through ATHENA software  based on the numerical
algorithms of IFEFFIT library~\cite{ravel2005athena}.
{Fig.~\ref{Fig:5} (a)} shows the fitted N K-edge XAS
spectra and the fitting involves an arc tangent step function
followed by four Gaussian envelopes corresponding to the features
`\textit{a}', `\textit{b}', `\textit{c}', `\textit{d}' and
detailed peak positions are listed in
{SM}~\cite{supplementaryinformation}. The N K-edge
features of TiN samples reveal that the doublet marked as
`\textit{a}' and `\textit{b}' near the threshold appears due to
the unoccupied hybridized molecular orbitals of TiN i.e. t$ _{2g}
$ (Ti 3d states overlapped with N 2p$ \pi $) and e$ _{g} $ (Ti 3d
states mixed with N 2p$ \sigma $) orbitals, respectively. Such
sharp intense peaks (`\textit{a}' and `\textit{b}') are
characteristic features of cubic rocksalt structure of
TiN~\cite{lazar2008nk}. The shoulder `\textit{c}' and broad feature
`\textit{d}' in the higher energy range are depicted as the
excitation to Ti 4sp hybridized with N 2p states. Feature
`\textit{e}' is due to intermixing of higher energy unoccupied
orbitals such as Ti 4p + N 2p-3p
states~\cite{pfluger1982electronic}. The octahedral crystal field
splitting 10Dq (E$ _{b} $ - E$ _{a} $) for TiN-1, TiN-5 and TiN-4
are found to be $ \approx $ 2.5 ($ \pm $0.1)\,eV whereas for TiN-6
sample it comes out to be $ \approx $ 2.6 ($ \pm $0.1)\,eV, which
matches well with the generally observed value of
2.5\,eV~\cite{chen1997nexafs,soriano1993interaction}. Further
proceeding towards the higher energy regime reveal that for
samples TiN-1 and TiN-4, the feature `\textit{c}' are alike having
steeper slopes in comparison to TiN-5 and TiN-6 samples. Such
smooth variation of feature `\textit{c}' have been witnessed for
strong (111) texture of TiN
films~\cite{esaka1997comparison,gupta2017early}. The variations can
be attributed to the long range ordering sensitivity of feature
`\textit{c}' due to higher scattering
effects~\cite{soriano1993interaction} and hence it is obvious that
highly close packed (111) textured films can suppress such
contribution.

To get an idea about the spin-orbit coupling, the Ti L-edge
features were also studied. Here, it is worth mentioning that Ti
L-edge consists of two features (L$ _{3} $ and L$ _{2} $) which
arises due to the transition of core electron from Ti 2p3/2$
\rightarrow $Ti 3d and Ti 2p1/2$ \rightarrow $Ti 3d
respectively~\cite{hu2001study}. Although, our obtained Ti L-edge
features are not consistent with the literature
reports~\cite{hu2001study,niesen2015titanium} and clearly confirms
the presence of surface oxidation which was in accordance with our
XRR fitting. From {Fig.~\ref{Fig:5} (b)} the four
features marked as `\textit{f}' (Ti 2p3/2$ \rightarrow
$O2p$\pi$+Ti3d), `\textit{g}' (Ti 2p3/2$ \rightarrow
$O2p$\sigma$+Ti3d), `\textit{h}' (Ti 2p1/2$ \rightarrow
$O2p$\pi$+Ti3d) and `\textit{i}' (Ti 2p1/2$ \rightarrow
$O2p$\sigma$+Ti3d) can be assigned to L$ _{3}$-t$ _{2g} $ L$ _{3}
$-e$ _{g} $, L$ _{2}$-t$ _{2g} $ and L$ _{2}$-e$ _{g} $
transitions which arise due to the octahedral crystal field
splitting~\cite{henderson2014x}. But surprisingly the 2p spin-orbit
coupling in all samples are found to be $ \approx $ 5.5\,eV which
is in well agreement with that of surface oxidation free TiN films
where a caping layer of Al was used to avoid surface
contaminations~\cite{hu2001study}. However, spectral modifications
is quite often in Ti L-edge spectra due to the multiplet effect
which plays a dominant role for early transition metal nitride
films like TiN~\cite{henderson2014x}. In essence, the Ti L-edge
measurements are in agreement with other measurement (e.g. XRR)
and the N K-edge spectra of TiN-5 and TiN-6 samples clearly
confirms the dominant (111) orientation as considered from XRD
measurements.

\section{Conclusion}
\label{4} In conclusion, the role of probing variable ion energy
without and with Ti interface on the resulting growth behavior of
TiN thin films with film microstructure, electrical behavior and electronic nature has
been studied systematically in details. The study reveals that
reduction in the ion energy results in denser and smoother films with appreciable reduction in the electrical resistivity, favoring the closely packed (111) preferred orientation which can
be interpreted in terms of surface kinetics of adatom mobility
pronounced for TiN thin films as calculated from SRIM. The {\Esp} which was initially $ \approx $ 35\,eV/atom at E$
_\mathrm{i} $ = 1\,keV reduced to $ \approx $ 27 eV/atom for E$
_\mathrm{i} $ = 0.5\,keV. Moreover, introduction of room temperature deposited Ti interface with 10\,nm thickness serves as a template for (111) texturing (also confirmed from our SXAS data)
of the films resulting in dense microstructure. Hence, it
leads to a mass density of 5.8\,g cm$ ^{-3} $ for TiN sample
deposited at E$ _\mathrm{i} $~=~0.5\,keV with Ti interface at room
temperature. It is worth mentioning that this is the highest
obtained density so far in TiN thin films to the best of our
knowledge and the reproducibility of the data was ensured several times through XRR measurements. In addition, these samples show substantially low resistivity ($ \approx $~90\,$ \mu $$ \Omega $ cm) value, which is among the lowest obtained values for room temperature deposited TiN films. Thus, it can be anticipated that IBS technique is one
of the preferred techniques where the appreciable modulation in
{\Esp} is especially effective for Ti interfaced TiN
thin films which can enhance the surface morphology of the
resulting ultradense crystal lattice with surface roughness down
to subnanometer ($<$ 1\,nm) regime and lowest resistivity values even at room temperature. Hence,
our findings depict pathways for enhanced surface morphology with low electrical transport response of
TiN thin films (at room temperature with no intentional
substrate bias or substrate heating) to an extent to avoid
interfacial diffusion probability of TiN thin films common in
thin film synthesis, usage in microelectronics and many more.

\section*{Acknowledgments}
Authors (SC and RG) are grateful to UGC-DAE CSR, Indore for
providing financial support through
CSR-IC-BL-62/CSR179-2016-17/843 project. Thanks are due to V. R.
Reddy and Anil Gome for XRR measurements, Rajeev Rawat and Dileep Kumar for the resistivity measurements, R. Venkatesh and Mohan
Gangrade for AFM measurements, Rakesh Sah and Avinash Wadikar for
SXAS measurement at BL-01, Indus 2, RRCAT, Indore, India. We also thank S.
Tokekar, A. K. Sinha, A. Banerjee and V. Ganesan for their kind
support and constant encouragements. SC is thankful to Yogesh and
Shailesh for the help provided in depositions.



\end{document}